\pdfoutput=1

\documentclass[11pt]{article}

\usepackage{acl}

\usepackage{times}
\usepackage{latexsym}

\usepackage[T1]{fontenc}

\usepackage[utf8]{inputenc}

\usepackage{microtype}

\usepackage{inconsolata}

\usepackage{graphicx,xcolor}  
\usepackage{pxrubrica}        
\usepackage{url}
\usepackage{booktabs}        
\usepackage{amsmath}         
\usepackage{multirow}        
\usepackage{ifthen}          
\usepackage{enumitem}        
\usepackage{comment}
\usepackage{algorithm}       
\usepackage{algpseudocode}

\usepackage{amsmath}
\usepackage[english]{babel}
\usepackage{hyperref}
\addto\captionsenglish{%
}
\addto\extrasenglish{%
}

\usepackage{listings}
\usepackage{xcolor}

\definecolor{codegreen}{rgb}{0,0.6,0}
\definecolor{codegray}{rgb}{0.5,0.5,0.5}
\definecolor{codepurple}{rgb}{0.58,0,0.82}
\definecolor{backcolour}{rgb}{0.95,0.95,0.92}

\lstdefinestyle{mystyle}{
    backgroundcolor=\color{backcolour},   
    commentstyle=\color{codegreen},
    keywordstyle=\color{magenta},
    numberstyle=\tiny\color{codegray},
    stringstyle=\color{codepurple},
    basicstyle=\ttfamily\scriptsize, 
    breakatwhitespace=false,         
    breaklines=true,                 
    captionpos=b,                    
    keepspaces=true,                 
    numbers=left,                    
    numbersep=5pt,                  
    showspaces=false,                
    showstringspaces=false,
    showtabs=false,                  
    tabsize=2
}

\usepackage{xcolor}          
\usepackage{pifont}
\newcommand{\pifontcheckmark}{\text{\ding{52}}}
\newcommand{\pifontcrossmark}{\text{\ding{56}}}
\newcommand{\good}{\textcolor{teal}{\pifontcheckmark}}
\newcommand{\bad}{\textcolor{magenta}{\pifontcrossmark}}

\title{Self-Organized Agents: A LLM Multi-Agent Framework toward \\Ultra Large-Scale Code Generation and Optimization}

\author{Yoichi Ishibashi  \\
  TsukushiAI \\
  \texttt{ishibashi.tsukushiai@gmail.com} \\\And
  Yoshimasa Nishimura \\
  TsukushiAI  \\
  \texttt{nishimura.tsukushiai@gmail.com} \\}

\begin{document}
\maketitle
\begin{abstract}
Recent advancements in automatic code generation using large language model (LLM) agent have brought us closer to the future of automated software development. However, existing single-agent approaches face limitations in generating and improving large-scale, complex codebases due to constraints in context length. To tackle this challenge, we propose \textbf{S}elf-\textbf{O}rganized multi-\textbf{A}gent framework (\textbf{SoA}), a novel multi-agent framework that enables the scalable and efficient generation and optimization of large-scale code. In SoA, self-organized agents operate independently to generate and modify code components while seamlessly collaborating to construct the overall codebase. A key feature of our framework is the automatic multiplication of agents based on problem complexity, allowing for dynamic scalability. This enables the overall code volume to be increased indefinitely according to the number of agents, while the amount of code managed by each agent remains constant. We evaluate SoA on the HumanEval benchmark and demonstrate that, compared to a single-agent system, each agent in SoA handles significantly less code, yet the overall generated code is substantially greater. Moreover, SoA surpasses the powerful single-agent baseline by 5\% in terms of Pass@1 accuracy.
\footnote{Our code will be available at \url{https://github.com/tsukushiAI/self-organized-agent}.}
\end{abstract}

\section{Introduction}
In recent years, research on agents using Large Language Models (LLMs)~\cite{DBLP:conf/nips/BrownMRSKDNSSAA20,DBLP:journals/corr/abs-2303-08774,DBLP:journals/corr/abs-2307-09288}, such as ReAct~\cite{DBLP:conf/iclr/YaoZYDSN023}, Reflexion~\cite{DBLP:conf/nips/ShinnCGNY23}, Toolformer~\cite{DBLP:conf/nips/SchickDDRLHZCS23}, and AutoGPT~\footnote{\url{https://github.com/Significant-Gravitas/}}, has been expanding the possibilities of automating human tasks. These advancements have particularly contributed to the rapid development of automatic code generation techniques in the field of automated application and tool development~\cite{DBLP:journals/corr/abs-2308-00352,DBLP:journals/corr/abs-2304-07590,DBLP:journals/corr/abs-2312-13010}. Compared to non-agent-based methods~\cite{DBLP:journals/corr/abs-2308-07124,DBLP:journals/corr/abs-2305-06161}, these research achievements have led to remarkable performance improvements in automatic code generation~\cite{DBLP:journals/corr/abs-2402-16906,DBLP:journals/corr/abs-2310-04406}.

\begin{figure}[t!]
\centering
\includegraphics[clip, width=8cm]{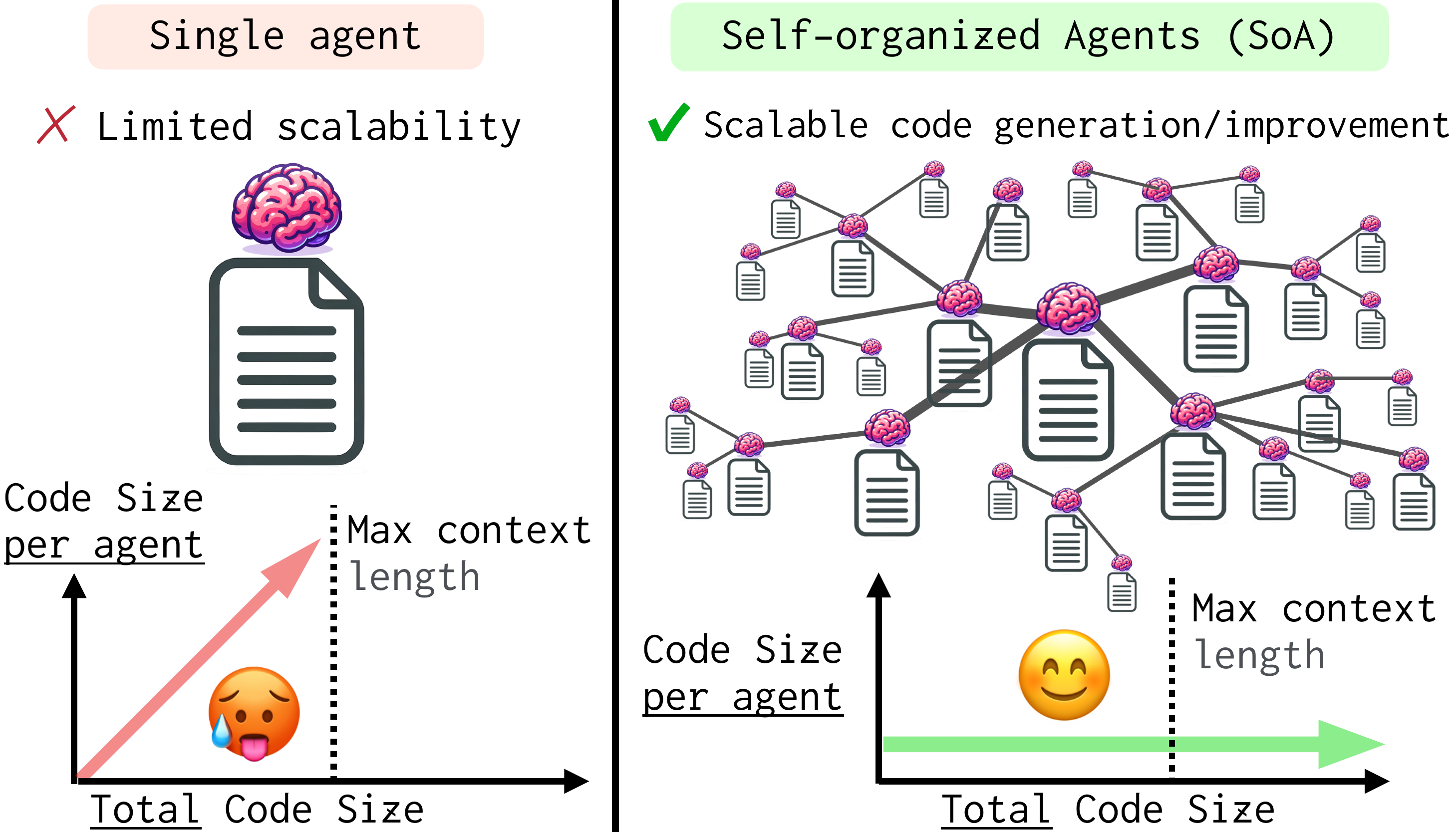}
\caption{
\textbf{Left (single agent):} A single agent is solely responsible for the entire implementation. As the codebase grows larger, the load increases for code generation, modification, and memory management, making it difficult to manage and develop. The larger the entire codebase becomes, the more it puts pressure on the context length during self-debugging, limiting the amount of code that can be managed.
\textbf{Right (SoA):} The implementation is distributed among multiple agents. The agents are independent; code generation, modification, and memory management are separated from other agents. Each agent manages only its own part, allowing it to focus on the implementation regardless of the complexity of the entire codebase. Furthermore, agents automatically multiply according to the complexity of the problem. This allows for the generation and modification of complex and large-scale code while maintaining a constant amount of code management/generation/modification per agent.
}
\label{fig:concept}
\end{figure}

Most recent research has focused on single-agent approaches for code generation. These single-agent code generation methods face limitations, especially in terms of scalability, when the implementation becomes complex and requires a large codebase. The main reason for this technical difficulty is that a single agent must manage the entire code generation process alone. For instance, implementing a machine learning algorithm involves several stages, such as data preprocessing, algorithm training, and result evaluation, which include many functions and classes. When these complex components are combined, the codebase inevitably becomes very large. 
However, there are limitations to the context length of LLMs, and as the number of input tokens increases, the inference performance decreases~\cite{DBLP:journals/corr/abs-2402-14848,DBLP:conf/emnlp/0002IEBL23,DBLP:journals/corr/abs-2311-04939}.
Consistently understanding and generating or modifying appropriate code for such an extensive codebase poses a significant challenge for a single agent in terms of comprehending and managing the context. Consequently, the single-agent approach struggles to efficiently generate and modify code as its complexity and size increase.

To tackle these challenges, we propose a self-organized multi agent framework that can automatically generate and modify large-scale code (\autoref{fig:concept}). \emph{Self-organization}~\cite{ashby1947principles} is a phenomenon in which living organisms or matter create an orderly, large structure as a result of their individual autonomous behaviors, despite lacking the ability to oversee the entire system.
In our framework, self-organized agents, each responsible for different code parts or tasks, independently generate and modify code. With the self-organization of agents, a single agent no longer needs to comprehend the entire codebase, making it possible to scale up large-scale code simply by increasing the number of agents.
Another feature of our framework is that agents automatically multiply according to the complexity of the problem, allowing the overall codebase to expand while keeping the amount of code handled by each agent constant.
These features enable the dynamic and flexible generation and modification of large-scale code, which was impossible with the traditional single-agent approach.

In our experiments, we evaluated the performance of this framework using HumanEval~\cite{DBLP:journals/corr/abs-2107-03374}, a benchmark for code generation. The results show that our self-organized multi-agent framework outperformed Reflexion~\cite{DBLP:conf/nips/ShinnCGNY23}, an existing powerful code generation agent (\autoref{sec:mainres}), demonstrating the effectiveness of our approach in generating and modifying code.
Furthermore, through a detailed analysis of the experimental results, we revealed how agents automatically multiply according to the complexity of the problem, effectively scaling up the overall code volume while keeping the code generation per agent constant (\autoref{sec:analysis}).
These experimental results support the contribution of our framework, which overcomes the scalability issues faced by single-agent approaches and provides a solution capable of handling larger projects.

\section{Code Generation Task}
The code generation task involves generating Python functions from docstrings~\cite{DBLP:journals/corr/abs-2107-03374}. In this task, an agent is given a docstring that defines the types of the function's inputs and expected outputs, as well as the specific requirements that the function should meet. The agent is then required to generate the code for a function that fulfills the specified functionality.
The generated code is verified for accuracy using unit tests, and the quality of the code is evaluated based on its ability to pass the test cases.
As with previous studies~\cite{DBLP:conf/nips/ShinnCGNY23,DBLP:journals/corr/abs-2402-16906,DBLP:journals/corr/abs-2310-04406}, we use the evaluation metric Pass@$1$~\cite{DBLP:journals/corr/abs-2107-03374}, where a problem is considered solved if any of the $k$ code samples pass all test cases.

\section{Self-organized Agent Framework}
\label{sec:soa}
Our Self-organized Agents (SoA) framework enables efficient implementation of large-scale and complex code by having self-organized agents independently generate and modify small-scale and simple code. In this section, we introduce the important components of SoA, namely the agents and the layers responsible for more abstract processing than the agents, and finally introduce the code generation and modification protocols in the SoA framework.

\subsection{Child Agent}
\label{sec:child}
\begin{figure*}[t!]
\centering
\includegraphics[clip, width=16cm]{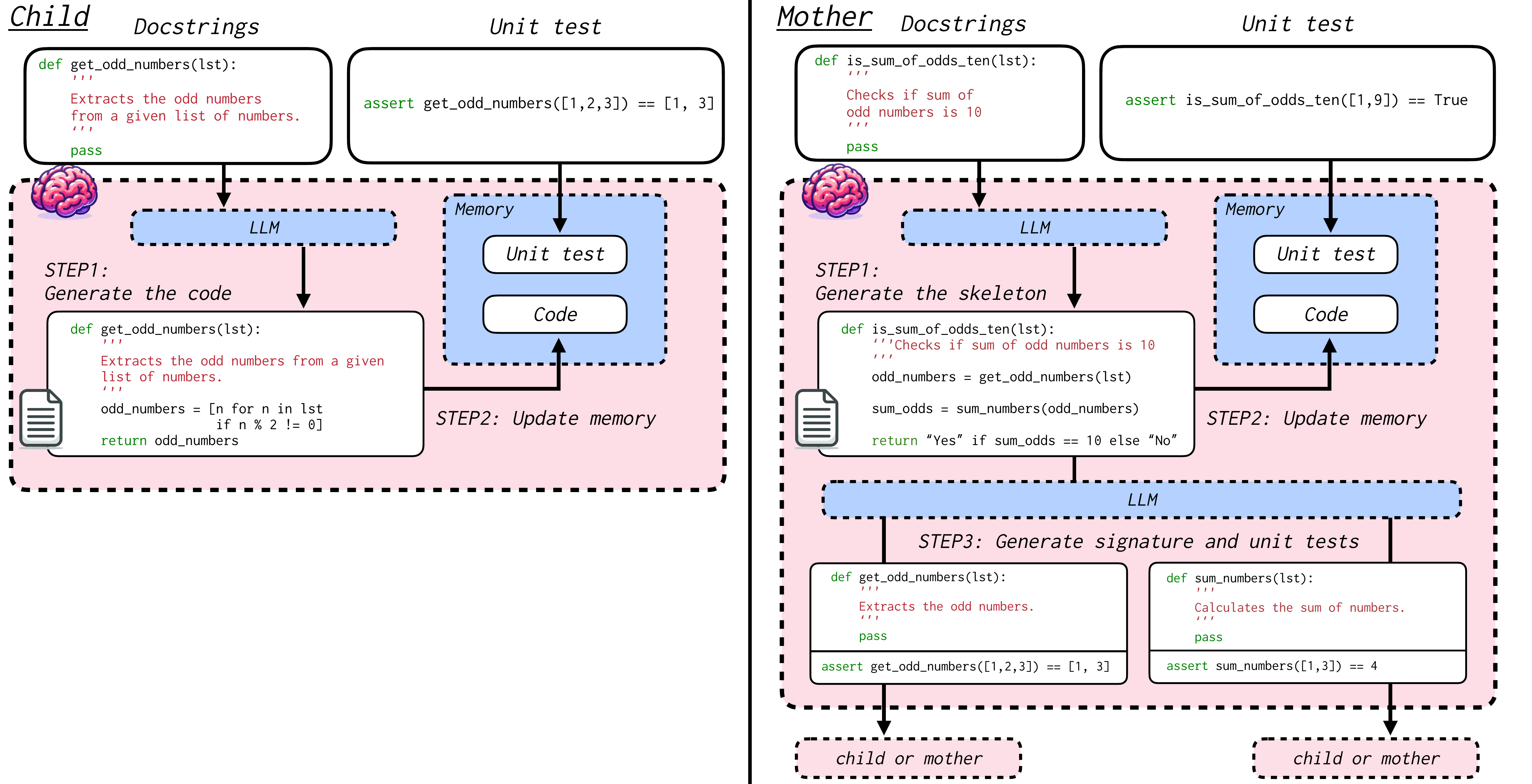}
\caption{
Overview of code generation. Child agents generate executable Python function from a given docstring. The Mother agent generates the skeleton of the function. The Mother spawns a new initialized agent (Child or Mother) and delegates unimplemented functions.
}
\label{fig:generation}
\end{figure*}

Child agents implement a given function based on its docstrings. As shown in \autoref{fig:generation}, this agent has a simple structure consisting of two elements: an LLM and memory. The LLM generates code from the given docstrings and modifies the code based on the results of unit tests. The memory stores the code generated by the agent itself and retrieves the latest code to be input to the LLM along with the unit test feedback during code modification. If an agent has these minimal specifications, it is possible to use an off-the-shelf agents (e.g., Reflexion) as a Child agent. We deliberately use a simple agent to verify the effectiveness of SoA in a simple setup.

\paragraph{Code Generation}
The main role of Child agents is to generate functions that meet the specifications based on the given function's docstrings. As shown in \autoref{fig:generation}, the agent follows the instructions to generate the rest of the function and complete it. The completed function implementation is stored in memory, and the unit tests for the function are also stored as they form the basis for future code modifications.

\paragraph{Code Modification: Empowering Child Agents with Self-Organization and Adaptability}
One of the most remarkable aspects of  agents in the SoA framework is their ability to autonomously improve their code based on the state of nearby agents . This process sets SoA apart from traditional agent approaches and showcases the power of self-organization in code modification. While existing agents like Reflexion~\cite{DBLP:conf/nips/ShinnCGNY23} rely solely on the results of unit tests, Child agents in SoA go beyond this limitation by independently observing the state of their mother agent, such as differences in modifications and feedback. By gathering this invaluable information from their surrounding environment, Child agents can adapt their behavior and make more informed decisions about code modification, even without explicit instructions.
The modifications and feedback generated by the Mother agent serve as an important source of information for the Child agents.
Armed with these insights, Child agents can more effectively modify their own code, contributing to the overall improvement of the codebase in a way that is both efficient and adaptive. \autoref{fig:self_debugging} illustrates  this process, which begins with the execution of unit tests and the retrieval of the latest implementation from memory. The Child agent then harnesses the power of the LLM to create a code modification proposal, seamlessly combining the information observed from the Mother agent with the test results and the latest implementation details. 
By storing the modified code in memory, Child agents create a feedback loop that continuously refines and improves the codebase over time. This iterative process, driven by the principles of self-organization and adaptability, enables SoA to tackle complex code modification tasks with  efficiency and effectiveness. As Child agents work in harmony with their Mother agent, they contribute to the creation of a more  optimized and large codebase.

\subsection{Mother Agent}
\label{sec:mother}
The Mother is an agent that generates new agents (Mother or Child).
Similar to Child agents, the Mother agent independently implements the specific Python function based on its given docstrings.
The Mother has memory, code generation capabilities, and self-debugging functions, as same as Child agents.
The unique feature of the Mother agent is its ability to generate multiple Child agents according to the complexity of the problem and delegate parts of the implementation to these agents.
This structure allows the Mother agent to focus on implementing abstract processes, while the Child agents generated by the Mother agent concentrate on implementing concrete processes. This division of labor enhances the overall efficiency and flexibility of the SoA framework.

\begin{figure}[t!]
\centering
\includegraphics[clip, width=7.7cm]{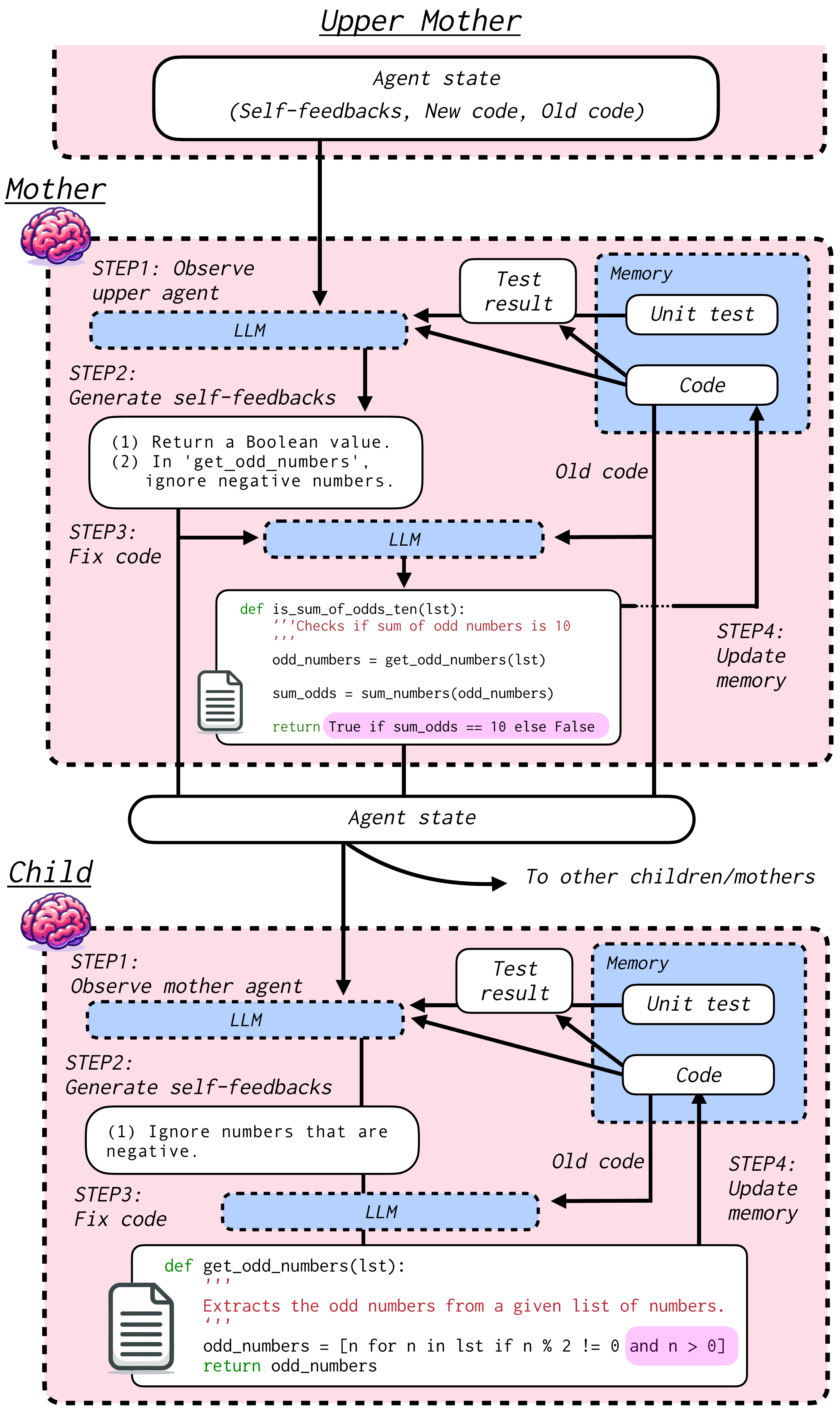}
\caption{
Overview of code modification. Agents (Mother/Child) observe the state of Mother (feedback, old code, and updated code) and use this information to improve the functions for which they are responsible. The state of the upper agent is used to modify code by lower agents within the hierarchy. This state propagation promotes collaboration and information sharing throughout the hierarchy, enabling efficient code modification.
}
\label{fig:self_debugging}
\end{figure}

\paragraph{Code Generation}
We explain the code generation process by the Mother agent using the implementation example of the \texttt{is\_sum\_of\_odds\_ten} function shown in \autoref{fig:generation}. The starting point is the function's docstrings and unit tests, which are memorized for reference in the later self-debugging phase. The first task of the Mother agent is to generate a skeleton of the implementation from the given docstrings, including subfunctions such as \texttt{get\_odd\_numbers} to extract odd numbers and \texttt{sum\_of\_numbers} to calculate their sum. The number and types of these subfunctions are automatically determined by the LLM based on the complexity of the problem.

It is important to note that these subfunctions are unimplemented, and the Mother agent does not directly implement them. Instead, it delegates the implementation of the subfunctions to other agents, allowing the Mother agent to focus on generating the skeleton and streamline its own code generation process. After the docstrings and unit tests for the subfunctions are generated, they are assigned to newly initialized agents for implementation. These agents proceed with the implementation of their respective functions without looking at the internals of the \texttt{is\_sum\_of\_odds\_ten} function implemented by the Mother agent. Since agents within the same Mother can work asynchronously, the overall code generation process is streamlined.

\paragraph{Code Modification}
The Mother's code modification is almost the same as the Child's code modification (\autoref{fig:self_debugging}). It observes information from the upper Mother and uses it to modify the functions it is responsible for. The only difference is that the feedback it generates and the code before and after modification are used by lower-level agents (Child or Mother).

\subsection{Self-organized Agent Process}
The Self-organized Agent (SoA) framework is a distributed framework in which multiple agents (including Mother agents and Child agents) repeatedly generate and modify functions. The core of this framework lies in the principle of self-organization, where each agent functions independently without the need to directly observe the entire codebase. The hierarchical combination of Mother agents and Child agents forms an agent network that effectively constructs a single large-scale codebase. In this hierarchical structure, Mother agents decompose complex problems into more manageable smaller problems by dividing tasks and delegating them to the agents they have generated. Although each agent is independent, the agents as a whole can work efficiently towards the implementation of a single function. Despite the fact that the amount of code each agent generates, modifies, and manages is always small, the number of agents scales, allowing the amount of code generated to be increased indefinitely according to the difficulty of the problem. Detailed algorithms are presented in Algorithm \autoref{alg:soa} in the appendix.

\begin{figure*}[t!]
\centering
\includegraphics[clip, width=16cm]{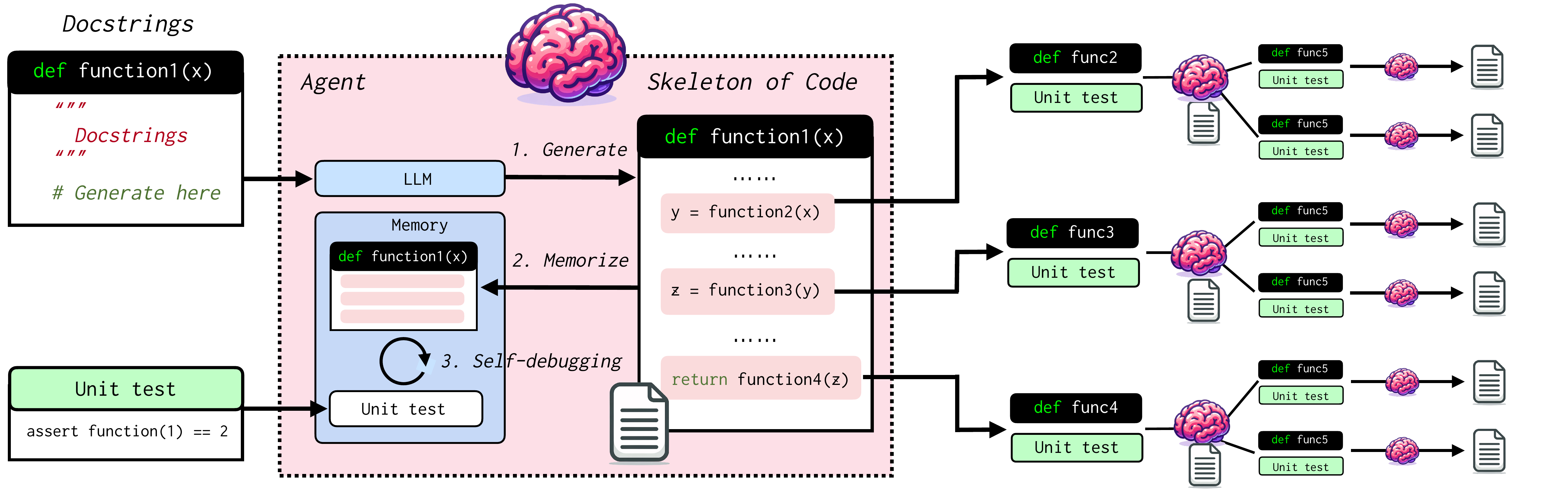}
\caption{
Overview of the SoA framework. Mother agents and Child agents hierarchically construct a network and perform function generation and modification. Mother agents delegate tasks to other Mother agents or Child agents, and each agent independently executes tasks while effectively implementing a single function as a whole.
}
\label{fig:overview}
\end{figure*}

\paragraph{Code Generation}
The code generation process in the SoA framework begins with the function's docstrings and unit tests. In the initial stage, there is only one initialized Mother agent, which is the root of the tree structure. Based on the input docstrings and unit tests, it generates docstrings and unit tests for subtasks and passes them to other agents it generates (see \autoref{sec:mother}). If the tree structure reaches a predetermined depth, the tasks are passed to Child agents; otherwise, they are passed to newly generated Mother agents. By repeatedly proliferating and increasing the number of agents until the last agent, it is possible to generate large-scale code while keeping the amount of code managed by individual agents constant.

\paragraph{Code Modification}
Once code generation is complete, the process transitions to the code modification phase. First, the implementations of all agents are combined to create the final implementation. This final implementation is evaluated using the unit tests provided to the root Mother, and feedback is generated from the results. Since there are no agents higher than this root Mother, information from higher-level agents as shown in \autoref{fig:self_debugging} is not used. The modification process starts based on this feedback and propagates information from the root Mother agent to the Child agents. Each agent updates its implementation based on the received feedback, generates new feedback, and transmits it to lower-level agents (see \autoref{sec:mother}). Finally, the Child agents update their own implementations, and the process terminates (see \autoref{sec:child}). This series of processes is repeated until a predetermined maximum number of iterations is reached.

\section{Experiments}
\paragraph{LLM Selection}
We used GPT3.5-turbo\footnote{\texttt{gpt3.5-turbo-1106}} for code generation and feedback generation.\footnote{GPT-4 was not selected due to the high experimental cost required.}


\paragraph{Baselines}
We compare SoA with several state-of-the-art code generation methods including AlphaCode~\cite{DBLP:journals/corr/abs-2203-07814}, Incoder~\cite{DBLP:conf/iclr/FriedAL0WSZYZL23}, Codex~\cite{DBLP:journals/corr/abs-2107-03374}, CoT~\cite{DBLP:conf/nips/Wei0SBIXCLZ22}, and Gemini Pro~\cite{DBLP:journals/corr/abs-2312-11805}. 
Additionally, we evaluate the performance of various GPT-3.5-based agents, such as  ChatGPT, Self-Edit~\cite{DBLP:conf/acl/ZhangLLLJ23}, and Reflexion~\cite{DBLP:conf/nips/ShinnCGNY23}. These baselines are chosen to represent a diverse range of approaches, including single-agent and multi-agent systems, as well as those with and without self-debugging capabilities. 

\paragraph{Agent Configuration}
To evaluate the effectiveness of the SoA framework, we selected the Reflexion agent as a baseline. Reflexion iteratively modifies code based on the given docstrings and automatically generated unit tests until it reaches the maximum number of iterations or passes the unit tests. The main difference between Reflexion and SoA is that Reflexion is composed of a single agent, while SoA is composed of self-organized multiple agents. 
In the SoA configuration, we set the maximum number of iterations for the learning loop to 8 and the maximum tree depth to 2.
Additionally, following \cite{DBLP:conf/nips/ShinnCGNY23}, we provided a few-shot trajectory to the LLM.

\paragraph{Data and Tasks}
To evaluate the performance of automatic code generation, we used the HumanEval \cite{DBLP:journals/corr/abs-2107-03374} benchmark. HumanEval is a set that includes diverse programming problems designed to measure the functional correctness of generated code. We used the Python language set for evaluation and followed the evaluation methodology of Reflexion \cite{DBLP:conf/nips/ShinnCGNY23}. In this process, multiple test cases are created for each generated code, and $n$ test cases are randomly selected to construct a test suite. This test suite is used to verify whether the generated code functions correctly. We set 6 unit tests for Reflexion and 1 unit test for SoA.

\begin{table}[t]
\setlength{\tabcolsep}{0.6mm} 
\small
\centering
\begin{tabular}{llccc}
\toprule
\multicolumn{2}{l}{\textbf{Method}} & \textbf{SD} & \textbf{SO} & \textbf{Pass@1} \\
\midrule
\multicolumn{2}{l}{AlphaCode}~\cite{DBLP:journals/corr/abs-2203-07814} & \bad  & \bad  & 17.1 \\
\multicolumn{2}{l}{Incoder}~\cite{DBLP:conf/iclr/FriedAL0WSZYZL23} & \bad  & \bad  & 15.2 \\
\multicolumn{2}{l}{Codex}~\cite{DBLP:journals/corr/abs-2107-03374} & \bad  & \bad  & 47.0 \\
\multicolumn{2}{l}{Gemini Pro}~\cite{DBLP:journals/corr/abs-2312-11805} & \bad  & \bad  & 67.7 \\
\midrule
\multirow{5}{*}{GPT-3.5} & \quad CoT~\cite{DBLP:conf/nips/Wei0SBIXCLZ22} & \bad  & \bad & 44.6 \\
& \quad ChatGPT & \bad & \bad  & 57.3 \\
& \quad Self-Edit~\cite{DBLP:conf/acl/ZhangLLLJ23} & \good & \bad & 62.2 \\
& \quad  Reflexion~\cite{DBLP:conf/nips/ShinnCGNY23} & \good & \bad & 66.5 \\
& \quad SoA (ours) & \good & \good & \textbf{71.4} \\
\bottomrule
\end{tabular}
\caption{Results of SoA and baselines on HumanEval. The score of ChatGPT is taken from \citet{DBLP:journals/corr/abs-2304-07590}. \textbf{SD} indicates whether the agent uses self-debugging with unit tests, while \textbf{SO} denotes whether the agent employs self-organized multi-agent collaboration.}
\label{tab:res_main}
\end{table}

\subsection{Main Results}
\label{sec:mainres}
\autoref{tab:res_main} compares the Pass@1 accuracy of the proposed method and the baseline. Comparing SoA with Reflexion, a strong baseline, SoA outperforms Reflexion by 5\% in Pass@1. Considering that each agent in SoA does not see the entire code, this is a surprising result. This result suggests that self-organized agents can generate code that functions well as a whole without needing to oversee the entire code, by independently implementing the functions assigned to them.

\subsection{Analysis}
\label{sec:analysis}
One of the most critical aspects of our study is  the efficiency of the self-organized multi-agent approach in large-scale code generation. To showcase the superior performance of SoA, we conducted a comprehensive comparative analysis between Reflexion, a state-of-the-art single-agent system, and our proposed multi-agent system. Using the HumanEval benchmark, we meticulously examined the overall scale of the code generated by both systems and the amount of code each agent independently generated and memorized. To ensure a fair comparison, we removed comments and docstrings from the HumanEval results and focused on the number of characters and tokens of pure code.

\begin{figure}[t]
\centering
\includegraphics[clip, width=8cm]{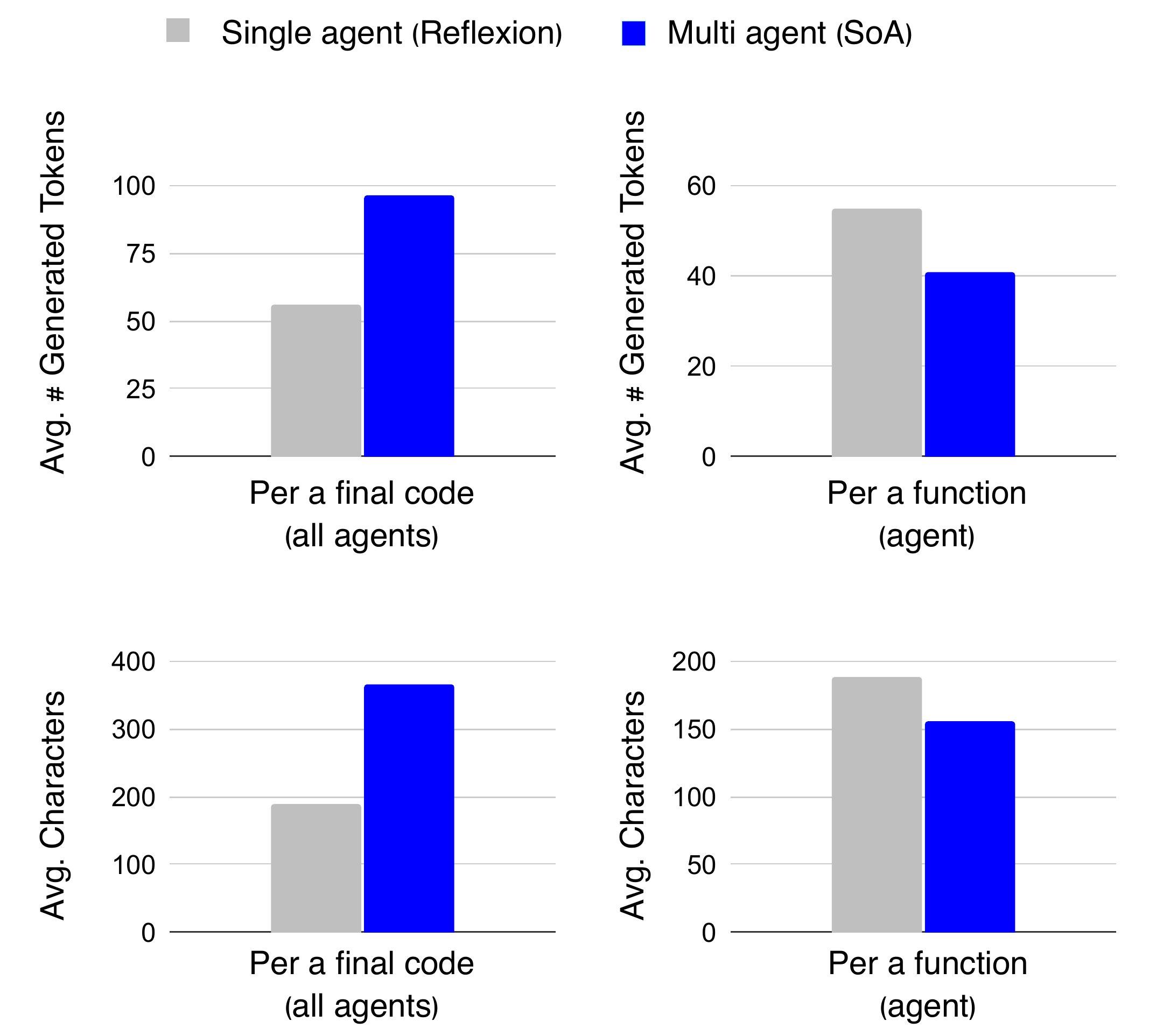}
\caption{
Comparison of code generation amount between SoA (mulit-agent) and Reflexion (single agent).
}
\label{fig:code_count}
\end{figure}

\autoref{fig:code_count} presents a  visualization of the average amount of code generated by SoA and Reflexion from the perspective of individual functions and all functions. In the context of HumanEval, which requires the implementation of a single function, SoA's code amount is calculated by summing the code generated by each agent, while Reflexion's code amount is based on a single function. The \emph{code amount per function} in SoA refers to the code generated by each individual agent, whereas in Reflexion, it is equivalent to the code amount of a single function. 
The results unequivocally demonstrate SoA's superiority over Reflexion in terms of the number of tokens per final code and the average number of characters per function. What is  remarkable is that despite each agent in SoA handling significantly fewer tokens/characters compared to the single agent in Reflexion, the overall output generated by SoA is substantially greater. This finding underscores the exceptional scalability of SoA, indicating its ability to handle increasingly complex tasks by seamlessly adding more agents to the system. Our results suggest that by increasing the depth of the agent hierarchy and introducing more Mother agents, SoA can generate even larger-scale code by efficiently distributing the workload among multiple agents. As the tree structure becomes deeper, the system exhibits an infinite scaling potential, enabling the generation of increasingly complex and extensive codebases while ensuring that each agent handles a manageable portion of the code. Each agent can maintain a manageable amount of code while theoretically allowing for an indefinite increase in the overall code generation capacity.

This distributed approach empowers SoA to significantly scale up its ability to tackle large-scale and complex coding tasks with remarkable efficiency and high quality, far surpassing the limitations encountered by single-agent systems like Reflexion, where a sole agent is responsible for managing and generating the entire codebase.

\section{Related Work}

\paragraph{LLM Agents}

Recent advancements in LLM agents, such as ReAct~\cite{DBLP:conf/iclr/YaoZYDSN023}, Reflexion~\cite{DBLP:conf/nips/ShinnCGNY23}, Toolformer~\cite{DBLP:conf/nips/SchickDDRLHZCS23}, and Self-Refine~\cite{DBLP:conf/nips/MadaanTGHGW0DPY23}, have primarily focused on single-agent approaches, where one agent is responsible for both generation and modification tasks. Among these, Reflexion~\cite{DBLP:conf/nips/ShinnCGNY23} has gained significant attention in the field of code generation due to its outstanding performance. However, despite their strengths, these single-agent approaches face inherent limitations when it comes to generating and modifying large-scale codebases. To address these limitations and push the boundaries of what is possible with LLM agents, we propose SoA, a novel multi-agent framework that harnesses the power of self-organization and collaboration. While we intentionally adopted simple agents for SoA in this work, our framework is flexible enough to incorporate more sophisticated and powerful methods~\cite{DBLP:journals/corr/abs-2402-16906,DBLP:journals/corr/abs-2310-04406} and other state-of-the-art LLMs~\footnote{\url{https://claude.ai/}}, further enhancing its potential for large-scale code generation and modification.

\paragraph{Multi-Agent Collaboration for Software Development}
In recent years, several multi-agent-based approaches have emerged as promising solutions for software development, such as MetaGPT~\cite{DBLP:journals/corr/abs-2308-00352}, ChatDev~\cite{DBLP:journals/corr/abs-2307-07924}, Self-collaboration~\cite{DBLP:journals/corr/abs-2304-07590}, and AgentCoder~\cite{DBLP:journals/corr/abs-2312-13010}. These methods typically personify agents and assign them specific names or occupational roles, such as programmers, project managers, or QA engineers, to allocate tasks. While this approach has shown promise, our method takes a different and more flexible approach. Instead of assigning fixed occupational roles, we subdivide agent capabilities based on \emph{code functionality}, allowing each agent to demonstrate its expertise without being constrained by predefined roles. This fine-grained task allocation enables more flexible problem-solving and adaptation to the complexity of the software development process. Moreover, by incorporating the concepts of self-organization and self-proliferation, our agents can dynamically scale up the overall code volume based on the difficulty of the problem at hand, providing a highly adaptable and efficient framework for large-scale code generation and modification.

\paragraph{Macro vs. Micro Perspectives}
While both multi-agent-based methods~\cite{DBLP:journals/corr/abs-2308-00352,DBLP:journals/corr/abs-2307-07924,DBLP:journals/corr/abs-2304-07590,DBLP:journals/corr/abs-2312-13010} and our proposed SoA framework share the common goal of automating software development, they address different technical aspects of the process. Existing multi-agent methods primarily focus on optimizing the macro structure of software development, such as project management and task allocation. In contrast, our method takes a more micro-level perspective, focusing on the elemental technologies of code generation and modification. These approaches are not mutually exclusive but rather complementary, offering a more comprehensive solution to the challenges faced in automatic software development. By combining the strengths of both macro and micro-level approaches, we can create a powerful and holistic framework that efficiently handles the complexities of large-scale code generation and modification.

\paragraph{Prompt Engineering}
Tree-of-Thought (ToT)~\cite{DBLP:conf/nips/YaoYZS00N23} and Skeleton of Thought (SoT)~\cite{DBLP:journals/corr/abs-2307-15337} are prompt engineering techniques that utilize tree-like structures. ToT represents reasoning steps as nodes to explore correct reasoning paths, while SoT generates a skeleton of the answer and completes the contents in parallel to decrease generation latency. In contrast, SoA uses a tree structure with agents as nodes, focusing on their collaboration and self-organization to generate and modify code efficiently. 

\section{Conclusion}
In this work, we introduced Self-organized Agents (SoA), a novel multi-agent framework for efficient and scalable automatic code generation and optimization using large language models (LLMs). SoA addresses the limitations of single-agent approaches in handling large-scale, complex codebases by leveraging the power of self-organization and distributed code generation. In SoA, self-organized agents operate independently to generate and modify code components while seamlessly collaborating to construct the overall codebase. A key feature of our framework is the automatic multiplication of agents based on problem complexity, allowing for dynamic scalability and enabling the overall code volume to be increased indefinitely according to the number of agents, while the amount of code managed by each agent remains constant.

We evaluated SoA on the HumanEval benchmark and demonstrated its superior performance compared to Reflexion, a state-of-the-art single-agent system, with SoA achieving a 5\% improvement in terms of Pass@1 accuracy. Furthermore, our in-depth analysis revealed SoA's remarkable scalability, as each agent in SoA handles significantly less code compared to the single-agent baseline, yet the overall generated code is substantially greater. These results highlight the effectiveness of SoA in generating and optimizing large-scale code efficiently and with high quality.

However, it is essential to acknowledge the limitations of the current implementation of SoA. The framework's performance may be affected by the choice of LLM and the quality of the generated unit tests. Additionally, SoA has been evaluated on a limited set of programming tasks, and its effectiveness in handling more complex, real-world software development projects remains to be investigated. Furthermore, the communication and collaboration mechanisms among agents in SoA can be further optimized to improve efficiency and fault tolerance.

Despite these limitations, we believe that the SoA framework has significant potential for future research and development in the field of automatic software development. 

\bibliography{custom}

\appendix

\section{Pseudocode}
\label{sec:appendix}


\begin{algorithm*}
\small
\caption{Generate Code with Self-organized Agent Framework}
\label{alg:soa}
\begin{algorithmic}[1]
\Require $docstrings$: Docstrings for the function, $unit\_tests$: List of unit tests, $max\_depth$: Maximum depth of the agent hierarchy, $max\_iterations$: Maximum number of code modification iterations
\Ensure The final generated code

\\

\State Initialize the root Mother agent with $docstrings$ and $unit\_tests$

\\

\Function{GenerateAgent}{$agent$, $depth$, $subtask\_docstrings$, $subtask\_unit\_tests$}
    \If{$depth - 1$ = $max\_depth$}
        \State $next\_agent \gets$ new ChildAgent
    \Else
        \State $next\_agent \gets$ new MotherAgent
    \EndIf
    \State Assign $subtask\_docstrings$ and $unit\_tests$ to $next\_agent$
    \State \Call{Generate}{$next\_agent$, $depth + 1$}
\EndFunction

\\

\Function{Generate}{$agent$, $depth$}
    \If{$depth$ = 1} \Comment{Root Mother}
        \State $skeleton \gets$ Generate skeleton from $agent.docstrings$ and $agent.unit_tests$
        \State $agent.code \gets skeleton$
        \For{each $subtask\_docstrings$, $subtask\_unit\_tests$ in subtasks}
            \State \Call{GenerateAgent}{$agent$, $depth$, $subtask\_docstrings$, $subtask\_unit\_tests$}
        \EndFor
    \ElsIf{$depth$ = $max\_depth$} \Comment{Child}
        \State Generate code for $agent.subtask\_docstrings$ and $agent.subtask\_unit\_tests$
        \State $agent.code \gets$ generated code
    \Else \Comment{Mother}
        \State Generate code for $agent.subtask\_docstrings$ and $agent.subtask\_unit\_tests$
        \State $agent.code \gets$ generated code
        \For{each $subtask\_docstrings$, $subtask\_unit\_tests$ in subtasks}
            \State \Call{GenerateAgent}{$agent$, $depth$, $subtask\_docstrings$, $subtask\_unit\_tests$}
        \EndFor
    \EndIf 
\EndFunction 

\\

\Function{Modify}{$agent$, $test\_result$, $upper\_agent\_observation$}
    \State Generate feedback for $agent$ based on $test\_result$ and $upper\_agent\_observation$
    \State Update $agent$'s code based on feedback
    \For{each $subagent$ in $agent.subagents$}
        \State Evaluate $subagent.code$ using $subagent.unit\_tests$ to get $subagent\_test\_result$
        \State \Call{Modify}{$subagent$, $subagent\_test\_result$, feedback and code changes}
    \EndFor
\EndFunction 

\\

\State Start code generation with \Call{Generate}{$root\_mother$, 1}

\\

\For{each iteration in $max\_iterations$}
    \State Combine implementations from all agents to create $final\_implementation$
    \State Evaluate $final\_implementation$ using $unit\_tests$ to get $test\_result$
    \State Modify the code starting from $root\_mother$ with \Call{Modify}{$root\_mother$, $test\_result$, None}
\EndFor 

\\

\State \Return The final implementation combined from all agents

\end{algorithmic}
\end{algorithm*}

\end{document}